# Revealing Polar Walls in $CsPbBr_3$: Herringbone Structure and Formation Pathway

Weilun Li[1], Qimu Yuan[2], Michael B. Johnston[2], Joanne Etheridge[1,3,4*]

[1] School of Physics and Astronomy, Monash University, Victoria, 3800, Australia

[2] Department of Physics, University of Oxford, Clarendon Laboratory, Oxford OX1 3PU, United Kingdom

[3] Monash Centre for Electron Microscopy, Monash University, Victoria, 3800, Australia

[4] Department of Materials Science and Engineering, Monash University, Victoria, 3800, Australia

## Abstract

Domain walls in ferroic materials can host nanoscale functionalities absent from the periodic crystal, offering new opportunities to control electronic transport and polarization in semiconductor devices. Metal halide perovskites demonstrate important photophysical properties; however, their ferroic properties remain unclear. Here we discover a hierarchical herringbone network of polar walls within the non-polar halide perovskite $CsPbBr_3$ comprising ferroelastic twin walls and antiphase walls. Atomic-scale imaging reveals that both wall types comprise nanoscale regions whose inversion-symmetry is broken compared to the surrounding bulk lattice. In-situ heating experiments show that these interfaces form sequentially during symmetry-lowering phase transitions, revealing a different formation pathway compared with 'conventional' oxide perovskites. These polar nano-walls, together with their enclosed topology, may impact carrier transport. This reveals a previously unrecognized structural motif in $CsPbBr_3$, providing new insights into their photophysical and optoelectronic behavior.

## Introduction

Metal halide perovskites (MHPs) have emerged as leading semiconductor materials for next-generation optoelectronic technologies, including solar cells[1], light-emitting diodes[2], lasers[3] and photodetectors[4]. These materials exhibit long photocarrier diffusion lengths and slow recombination kinetics despite relatively high densities of intrinsic crystalline defects[5]. Although these exceptional photophysical properties have enabled rapid advances in device performance, their microscopic structural origins, however, remain incompletely understood[6].

Domain walls are two-dimensional topological defects that separate regions with distinct order parameters, such as strain or polarization[7]. In perovskite materials, the flexible corner-sharing octahedral framework readily accommodates lattice distortions, octahedral tilting, and symmetry breaking within locally-confined regions. These confined interfaces can therefore exhibit functionalities that do not present in the bulk crystal, including polarization in nominally non-polar lattices and enhanced electronic conductivity[8–10]. Such behavior has led to the hypothesis that domain walls may facilitate charge separation and carrier transport in MHPs[11–14].

In widely studied MHPs, such as $MAPbI_3$ ($MA=CH_3NH_3$) and $CsPbBr_3$, intragrain domains are commonly observed and are generally attributed to ferroelasticity arising from strain relaxation during structural phase transitions[15–20]. Although the ferroelectric nature of these materials remains debated, most studies assign non-polar space groups at room temperature, tetragonal *I4/mcm* for $MAPbI_3$ and orthorhombic *Pbnm* for $CsPbBr_3$[10,17,21]. Ferroelastic domain walls have nevertheless been proposed to promote electron-hole separation and serve as efficient pathways for photocarrier transport[11–14]. These effects are often attributed to strain gradients at the walls that locally break inversion symmetry and generate polarization through flexoelectric coupling.

Despite growing interest, no direct atomic-scale evidence has been provided. Most existing evidence has been inferred from mesoscopic imaging, diffraction or macroscopic measurements, while direct atomic-scale observations remain scarce[22].

Another important unresolved issue concerns the relationship between domain walls and structural phase transitions. Because ferroelastic domains originate from symmetry-lowering transformations, the structure and stability of domain walls are expected to evolve with changes in crystal symmetry[16,17,19,23]. However, this evolution has rarely been observed directly in halide perovskites, limiting our understanding of how domain-wall microstructure forms and how their formation can be engineered.

Here we address these questions in vacuum-deposited $CsPbBr_3$ thin films, an important platform for scalable and stable perovskite optoelectronics and photonics[3], using low-dose transmission electron microscopy (TEM).

## Results

### Herringbone ferroelastic domain

We first examine the microstructure of polycrystalline $CsPbBr_3$ thin films with a thickness of ~200 nm, comparable to typical optical gain media in laser cavities. Bright-field TEM reveals widespread lamellar domain structures within individual grains (**Fig. 1A**), consistent with previous diffraction evidence[10,20]. The domains appear as parallel stripes with alternating intensity across domain walls in grains aligned close to the <$1\bar{1}0$> zone axis. Notably, the domain widths of ~50-150 nm increase with grain size (**Fig.S1**), consistent with the scaling behavior for ferroelastic domains[24].

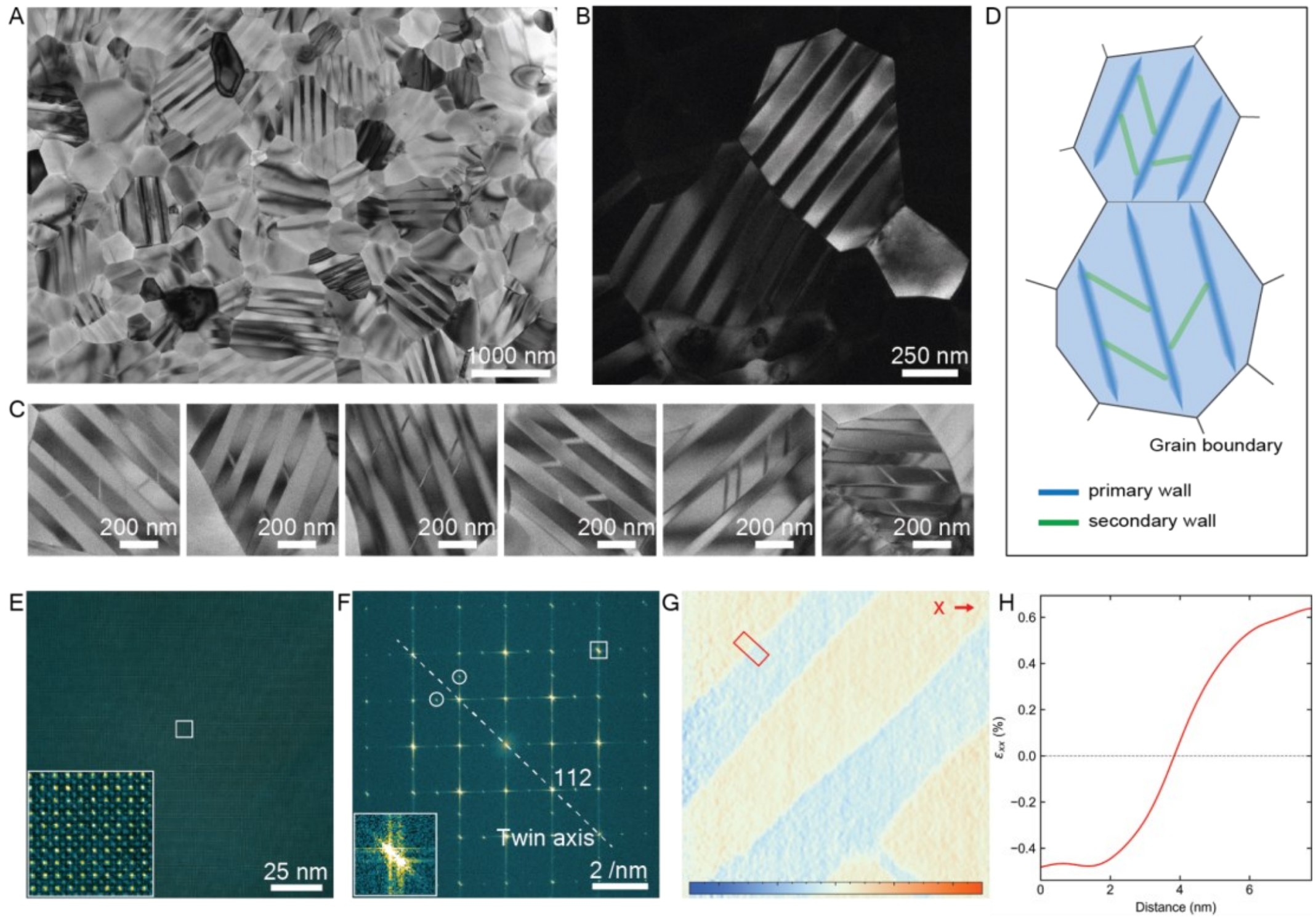

**Fig. 1. Herringbone domain microstructure in $CsPbBr_3$ thin films. (A)** Bright-field TEM image of a polycrystalline $CsPbBr_3$ film showing widespread lamellar domains within grains. **(B)** Dark-field TEM image highlighting alternating diffraction contrast across adjacent domains. **(C)** Higher magnification TEM images showing secondary domain walls intersecting the primary ferroelastic twins. **(D)** Schematic illustrating the hierarchical wall architecture. **(E)** Atomic-resolution scanning transmission electron microscopy-annular dark field (STEM-ADF) image acquired from a $[1\bar{1}0]$-oriented grain. **(F)** Fourier transform of the image in **(E)**. **(G)** Lattice parameter map in the x-direction measured by geometric phase analysis (**Fig. S3**). The color scale ranges from -5% to 5% (left to right). **(H)** Line profile from the region highlighted in **(G)** shows the lattice parameter change across the twin wall.

Dark-field TEM images further support the ferroelastic nature of these domains. Alternating bright-dark diffraction contrast is observed along the lamellae (**Fig. 1B**), indicating small crystallographic misorientations between neighboring domains, characteristic of ferroelastic twins. In particular, in some grains, the domains terminate at grain boundaries with needle-like morphologies, reflecting the minimization of elastic energy at the intersection between ferroelastic domains and grain boundaries.

Closer inspection reveals a second set of nanoscale walls embedded within the larger lamellar domains. These secondary walls are oriented at approximately 45° relative to the primary walls (**Fig. 1C**). Together, the two wall types form a hierarchical herringbone-like pattern across the grain (**Fig. 1D**).

To resolve the atomic structure of these walls, we examined thinner $CsPbBr_3$ films with a nominal thickness of ~35 nm, optimized for atomic-resolution TEM. The same herringbone domain pattern is consistently observed (**Fig. 1(E-H)**), confirming that it is intrinsic to the material regardless of film thickness. Fourier transform analysis of the atomic-resolution image (**Fig. 1F**) shows mirrored spots across the wall, confirming a twin relationship (**Fig. S2**). The swap in lattice parameter across the ferroelastic twin walls is illustrated by the measurement of lattice spacing by geometric phase analysis (**Fig. 1G**). Importantly, it also reveals a strain gradient at the ferroelastic twin wall (**Fig. 1H**).

The primary domain walls correspond to {112} ferroelastic twin planes, representing a ~90° (measured as 88.2°) twin relationship. In contrast, the secondary walls lie approximately along {110} or {002} planes and do not change the crystal orientation and strain state across the interface, indicating that they are structurally distinct from the ferroelastic twin walls (**Fig. S4**).

Together, these results reveal that $CsPbBr_3$ films host a previously unrecognized hierarchical wall architecture, in which {112} ferroelastic twin walls are interwoven with {002} and {110} walls, forming a herringbone network of walls within individual grains.

**Atomic structure of ferroelastic twin and antiphase walls**

We next quantified the position of atomic columns from atomic-resolution images to identify the structural origin of domain walls. In orthorhombic $CsPbBr_3$, the *Pbnm* structure (space group #62; equivalently *Pnma* under an alternative axis definition) produces antiparallel off-center displacements of Cs cations (**Fig. 2(A-C)**). We use this displacement to map local structural variations across the walls.

Ferroelastic twin walls (TWs) are identified by a 90° rotation of the Cs displacement pattern across the interface, consistent with the crystallographic relationship of {112} ferroelastic twin (**Fig. 2D**). Within the ferroelastic domains, we also observe a second type of wall that preserves the crystal orientation but reverses the phase of the Cs displacement pattern. For the {002} wall, the Cs displacement gradually decreases toward the interface, reaches a minimum at the wall, and reverses by 180° across it (**Fig. 2E**). This indicates a half unit-cell translation between neighboring domains, identifying the interface as an antiphase wall (APW), denoted as APW(**I**). A similar wall configuration occurs along {110} planes, denoted APW(**II**), where the Cs displacement lies head-to-head (**Fig. 2F**).

Both TWs and APWs show a local suppression of Cs displacement at the interfaces. The gradual reduction and reversal of this displacement order parameter indicates that the walls locally modify the *Pbnm* symmetry, allowing Cs cations to relax toward central positions. This reconstruction is dominated by Cs rather than Pb displacements relative to the centrosymmetric positions (**Fig. S5**). We note that minor measurement artefacts may arise from slight crystal misorientation (**Fig. S6**).

These observations demonstrate that the hierarchical wall architecture in $CsPbBr_3$ consists of three distinct domain walls: {112} TWs and {002} APW(**I**) and {110} APW(**II**) where the Cs displacement field is locally suppressed and switched across the wall.

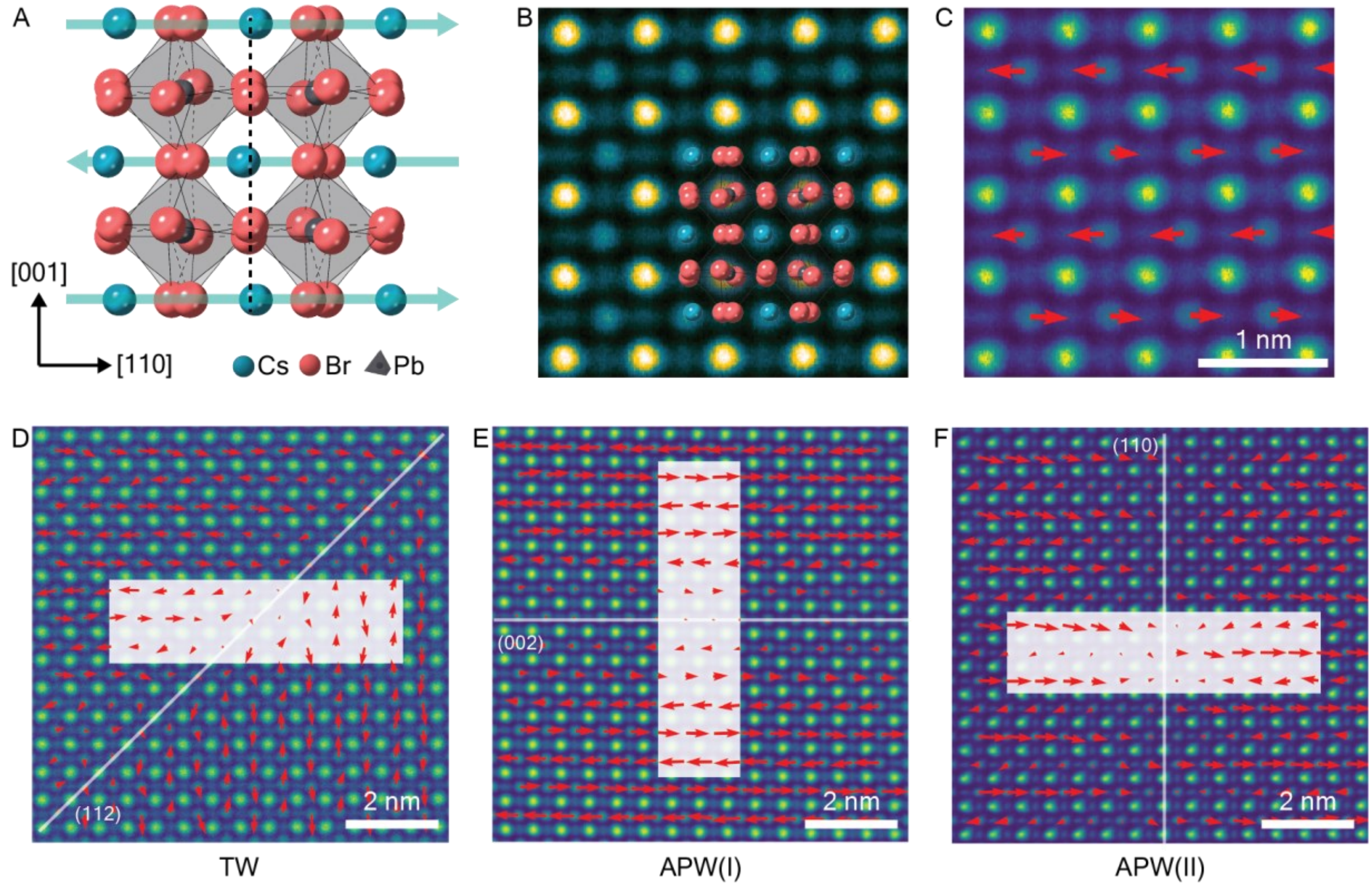

**Fig. 2. Atomic structure of ferroelastic twin walls and antiphase walls. (A)** Structural model of orthorhombic $CsPbBr_3$. Green arrows show antiparallel Cs off-center displacements. Dashed line shows the central positions of Cs cations. **(B)** Atomic-resolution STEM-ADF image overlaid with the structural model along the $[1\bar{1}0]$ zone axis. **(C)** Displacement map illustrating the Cs displacement relative to the center of neighboring Pb atomic columns. **(D-F)** Displacement maps across a **(D)** {112} ferroelastic twin wall; **(E)** {002} antiphase wall; **(F)** {110} antiphase wall.

### Coupled cation displacement and octahedral tilting

Octahedral tilting is another key structural order parameter controlling the lattice symmetry and properties of MHPs[25]. In orthorhombic $CsPbBr_3$, the $PbBr_6$ octahedra adopt an $a^-a^-c^+$ tilt pattern in Glazer notation[26]. The tilt components can be estimated from the shape anisotropy of Pb/Br atomic columns, which reflect the projected distortions of the octahedral framework (**Fig. S7**).

Our analysis reveals a strong correlation between the Cs displacement and octahedral tilting across the domain walls. Taking the APW(I) as an example, regions far from the wall show the expected *Pbnm* structure, with antipolar Cs displacements, accompanied by alternating octahedral tilts (**Fig. 3**). Toward the domain wall, the Cs displacement progressively decreases, while the octahedral tilt pattern relaxes. At the wall, the out-of-plane tilt component is largely suppressed, changing from $a^-a^-c^+$ to $a^0a^0c^+$. The projected structure is therefore consistent with a local symmetry approaching the tetragonal *P4/mbm* phase. Similar behavior is observed at the TW and APW(II) (**Fig. S8**). This coupling between A-site displacement and octahedral tilting governs the atomic reconstruction of the walls. The displacement gradient produces a residual Cs displacement of up to 6.5 pm between each pair of antipolar sites. This displacement can be converted into a projected polarization $P_s$ with the equation $P_s = \frac{1}{V}\sum \delta_i Z_i$, where $V$ is the volume of unit cell, $\delta_i$ is the displacement, and $Z_i$ is the Born effective charge of atom $i$. This yields an estimated polarization of 0.71 μC $cm^{-2}$ and an ideal unscreened field of 1.74 x $10^8$ V/m, using $Z_{Cs}$ = 1.31 and a dielectric constant $\epsilon$ = 5.70, obtained from density functional theory for locally distorted, polymorphous $CsPbBr_3$[27].

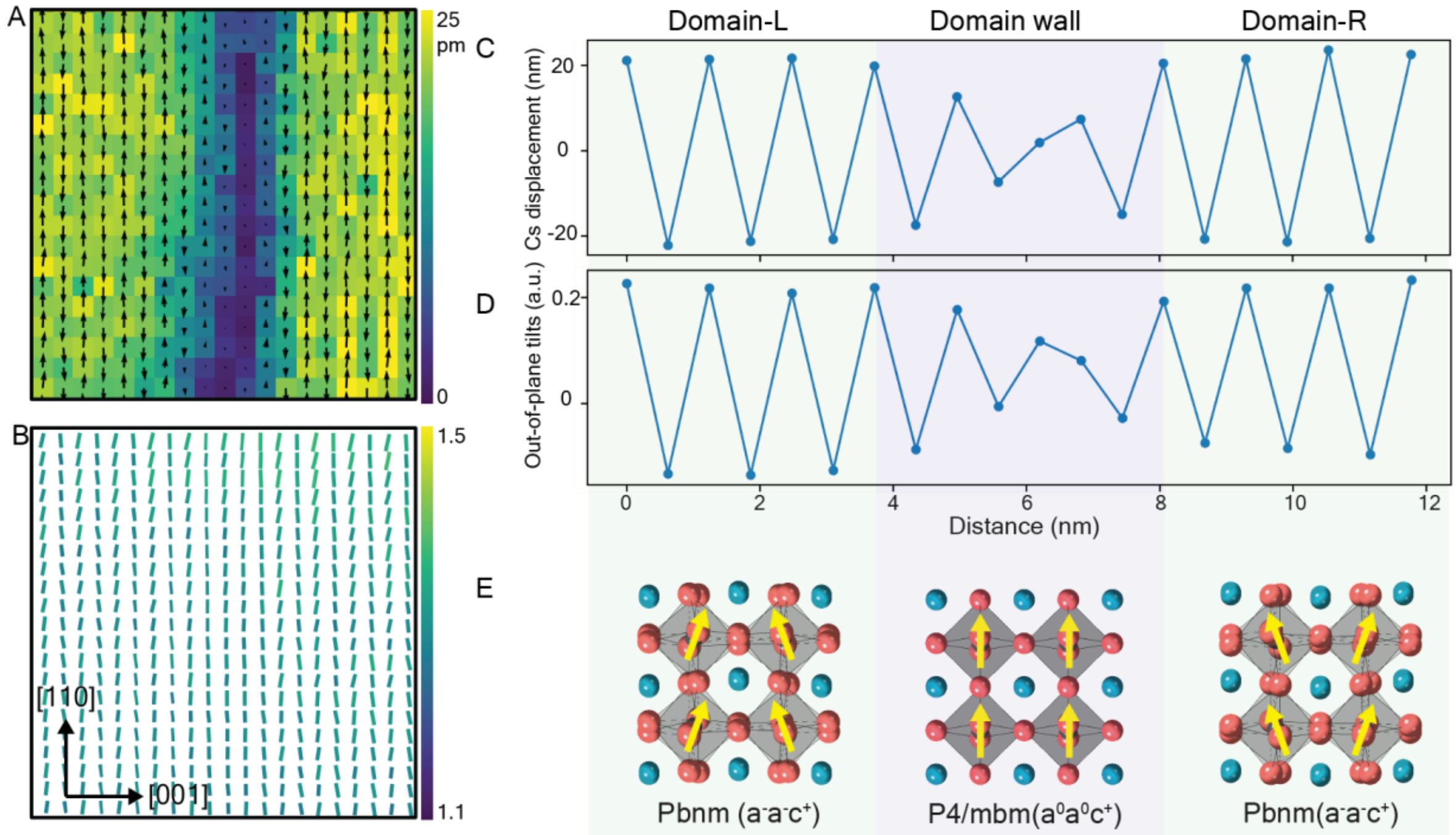

**Fig. 3. Correlated Cs displacement and octahedral tilts across an APW(I) in $CsPbBr_3$. (A)** Map of Cs displacement. **(B)** Ellipticity vector map measured from Pb/Br columns. Line direction indicates the orientation of the ellipse major axis. Line length and color indicate the magnitude of ellipticity (major axis/minor axis). **(C)** Line profile of the Cs displacement in the [110] direction. **(D)** Line profile of angle of the ellipticity vector relative to the [110] direction. **(E)** Schematic structural models illustrating the symmetry lowering across the wall.

### Evolution of domain walls across phase transitions

The crystal symmetry of MHPs is strongly temperature dependent. Bulk $CsPbBr_3$ adopts an orthorhombic *Pbnm* phase at room temperature, transforms to a tetragonal *P4/mbm* phase at ~88°C, and becomes cubic $Pm\bar{3}m$ phase above ~130°C[17,19,28]. To investigate how domain walls respond to these bulk symmetry changes, we performed in-situ heating and cooling experiments in the TEM while carefully minimizing the electron dose to avoid beam-induced degradation.

At room temperature, the polycrystalline film exhibits widespread lamellar domains within individual grains (**Fig. 4A**). Upon heating to 150 °C, corresponding to the cubic phase, these domains disappear (**Fig. 4B**). This observation indicates that the ferroelastic and antiphase domains originate from symmetry breaking associated with the lower-symmetry phases. When the specimen is cooled back to room temperature, the domain structures reappear (**Fig. 4C**), demonstrating that the domain formation process is reversible and governed by the structural phase transitions.

Close examination reveals that the two types of domain walls evolve differently during these transitions. TWs vanish and reappear during the cubic-tetragonal transition, whereas APWs emerge only during the tetragonal-orthorhombic transition (**Fig. 4D**). This sequential evolution suggests that the TWs form first as the lattice lowers from cubic to tetragonal symmetry, while APWs subsequently develop within the twin domains when the orthorhombic structure appears. Such a formation sequence naturally explains the hierarchical herringbone pattern observed at room temperature, where APWs are embedded within the ferroelastic domains.

During repeated heating and cooling cycles, the positions and orientations of domain walls are not strictly preserved. In some grains, the orientation of TWs flips by ~90°, indicating ferroelastic switching between equivalent crystallographic {112} planes. Statistical analysis of 121 grains shows that most grains (73 out of 121) retain their original domain orientation after the thermal cycle, while the remaining grains undergo 90° switching (**Fig. 4E**). This suggests there are no underlying structural defects that might pin the location of the twin walls, consistent with high crystal quality. Interestingly, grains that switch their domain orientation tend to cluster spatially, suggesting that ferroelastic switching is mechanically coupled between neighboring grains (**Fig. 4(F, G)**). This collective behavior indicates that long-range elastic interactions play an important role in governing the evolution of domain structures during phase transitions. Indeed, it is common to observe the twin wall orientation and domain width to align across a grain boundary (**Fig. S9**).

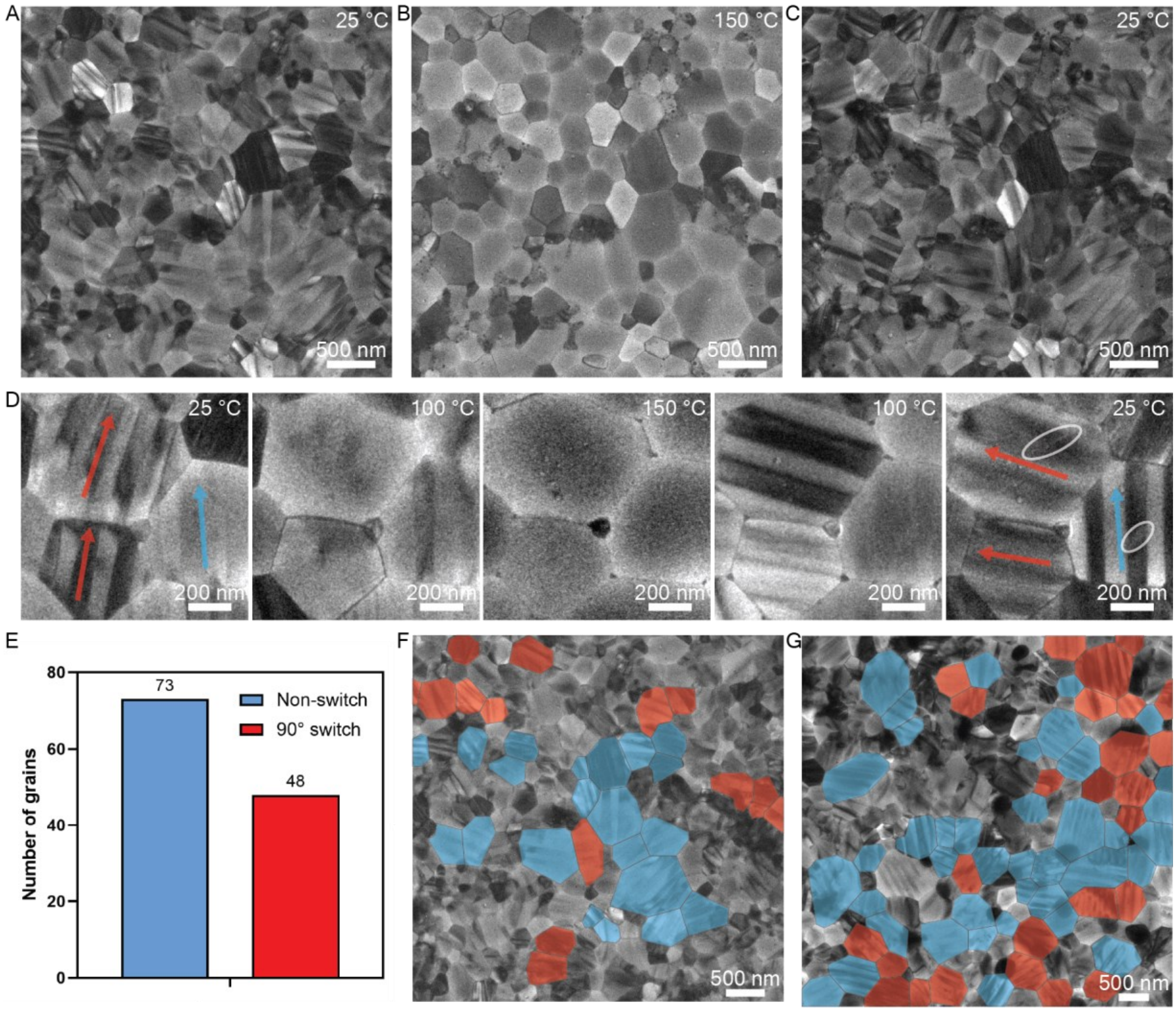

**Fig. 4. Evolution of domain structures across structural phase transitions in $CsPbBr_3$. (A)** TEM image of a $CsPbBr_3$ thin film at 25°C and the same region after heating to **(B)** 150 °C, followed by cooling back to **(C)** 25°C. **(D)** temperature-controlled experiment illustrating the evolution of TWs and APWs during heating and cooling. Arrows indicate ferroelastic domain orientations. Ovals mark the positions of APWs. **(E)** Statistic showing numbers of domains retain or switch their ferroelastic-domain orientation after a heating-cooling cycle. **(F, G)** TEM images highlighting grains that retain their original ferroelastic orientation (blue) and those that undergo 90° switching (red). Individual images at different temperatures are presented in **Fig. S10** and **Fig. S11**.

## Discussion

Our results reveal that $CsPbBr_3$ films contain a hierarchical herringbone network of ferroelastic twin walls (TWs) and antiphase walls (APWs). This herringbone structure has a fundamentally different formation pathway compared to conventional oxide perovskites. Unlike the herringbone pattern in ferroelectric oxide perovskites, such as $BaTiO_3$[29,30] and $PbTiO_3$[31–33], this network forms in a centrosymmetric halide perovskite where bulk ferroelectric polarization cannot exist. The microstructure therefore arises not from conventional 90° and 180° ferroelectric domains, but from coupled ferroelastic and antiphase domain formation.

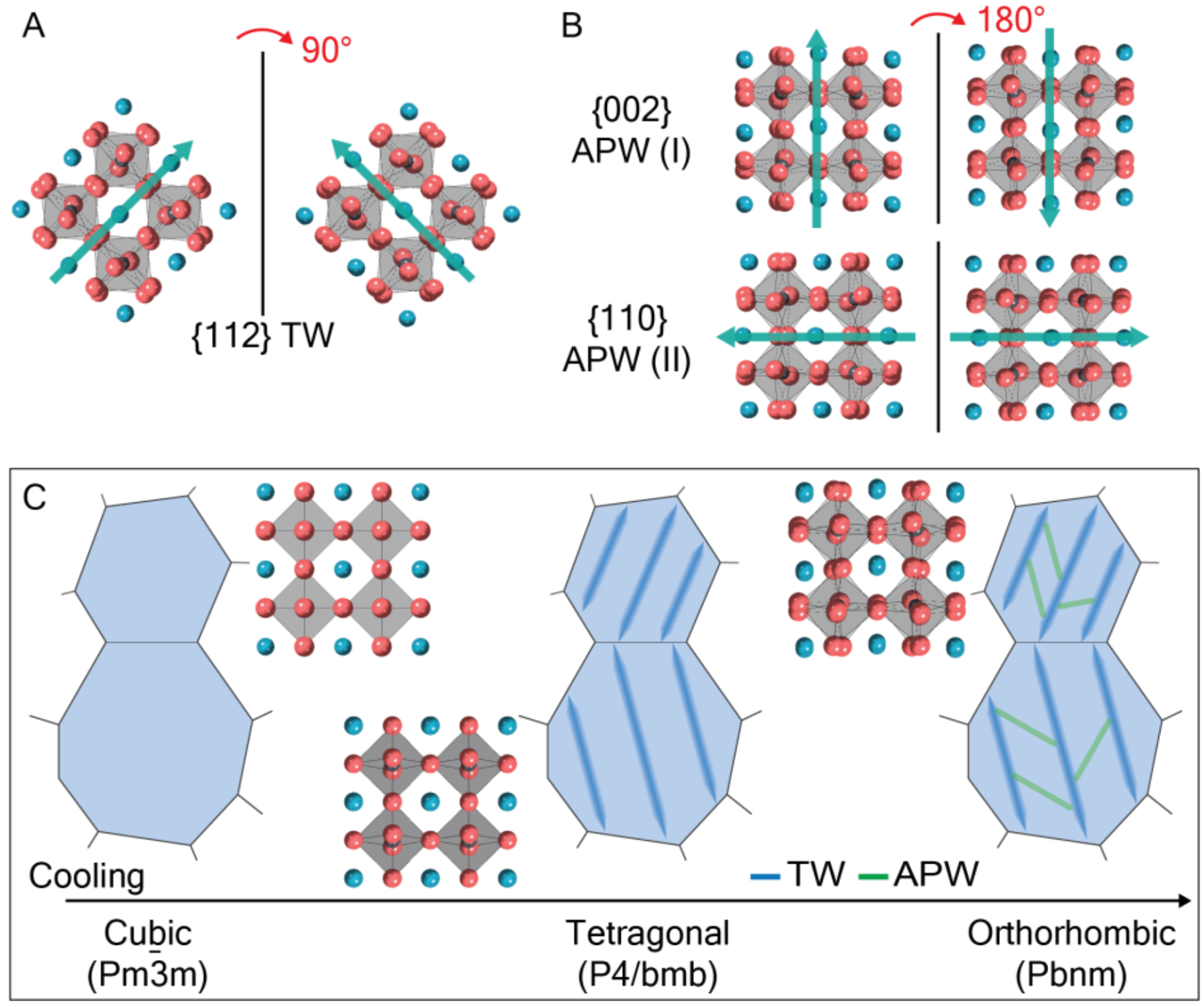

**Fig. 5. Schematic diagrams showing the atomic structure of domain walls in $CsPbBr_3$ films. (A)** Ferroelastic twin walls. Arrows indicate Cs displacement directions. **(B)** Antiphase walls. **(C)** Evolution pathway of domain walls with phase transitions.

Atomic-resolution imaging further shows that both the TWs and APWs host pronounced local structural changes. In the orthorhombic *Pbnm* phase, antipolar Cs displacements ($R_5^+$, following the notation of Miller and Love[34]) coupled with octahedral tilts ($R_4^+$ and $M_3^+$) stabilize the centrosymmetric lattice and suppress bulk ferroelectricity[35,36]. At the walls, the Cs displacement progressively decreases toward the interface, accompanied by relaxation of the octahedral tilts. The wall region therefore acts as a nanoscale structural transition zone in which the orthorhombic order parameters are reduced, consistent with a Landau–Ginzburg description of continuous order parameter variation across domain walls.

The antipolar atomic displacement gradient breaks the inversion symmetry constraint imposed by the bulk lattice, generating a locally polar structure that is otherwise forbidden in the centrosymmetric crystal. The domain walls in $CsPbBr_3$ locally disrupt the antipolar Cs displacement pattern. At both TWs and APWs, neighboring Cs columns exhibit unequal displacement amplitudes across the interface, preventing complete cancellation of the antipolar A-site distortions present in the bulk lattice. This mechanism of incomplete cancellation of antipolar A-site displacements has been used to induce ferroelectricity in engineered oxide superlattices, where alternating $1ABO/1A'BO_3$ layers break the symmetry that enforces antipolar compensation in the parent lattice[37]. By analogy, this imbalance in $CsPbBr_3$ likely produces a residual dipole moment and generates a localized electric field within the interface.

## Conclusions

In summary, we reveal the presence of a hierarchical herringbone domain structure with polar domain walls in polycrystalline $CsPbBr_3$ thin films and determine the atomic structure and formation pathway of these domain walls. Low-dose TEM reveals a network of {112} ferroelastic twin walls interwoven with {110} and {002} antiphase walls. In situ heating shows that these walls form sequentially during symmetry-lowering phase transitions, while atomic-resolution imaging reveals suppressed antipolar Cs displacements and relaxed octahedral tilts at the wall regions. The topologically connected array of walls may constrain transport pathways in optoelectronic devices, such as perovskite light-emitting diodes. Furthermore, all three types of domain wall have the potential to host a localized electric field due to the local symmetry breaking and incomplete cancellation of antipolar displacements at domain walls. Because the wall network is governed by phase-transition symmetry and kinetics, its density, orientation and connectivity should be tunable through composition, growth conditions, strain and thermal annealing, providing opportunities to engineer domain walls as a new structural element in halide perovskite optoelectronic and photonic devices.

## Methods

**$CsPbBr_3$ thin-film sample fabrication.** $CsPbBr_3$ samples were fabricated directly on TEM grids (Agar Scientific, Cu-400 mesh) through a previously reported vapor co-deposition method in our custom-built thermal evaporator[3]. Prior to all depositions, TEM grids were $O_2$-plasma treated for 0.3 minutes. The chamber was pumped down to a base pressure below $3 \times 10^{-6}$ mbar for all depositions, and precursors of CsBr (Alfa-Aesar, 99.9% metals base) and $PbBr_2$ (Alfa-Aeser, 99.998%, metals base) were used. An optimized nominal precursor flux ratio of CsBr:$PbBr_2$ = 1.5:1 was utilized[3] and a thickness of 35 nm was deposited. As-deposited films were immediately annealed at 200 °C in $N_2$ atmosphere for 2 minutes. For the sample transfer from Oxford, United Kingdom to Melbourne, Australia, the samples were stored and securely sealed inside a light-proof, stainless-steel transfer unit in $N_2$ at atmospheric pressure. Courier transfer was typically completed within 5 days.

**Transmission electron microscopy (TEM).** Scanning Transmission Electron Microscope - Annular Dark Field (STEM-ADF) images were taken using a Thermo Fisher Scientific Spectra φ Field Emission Gun Transmission Electron Microscopy (FEG-TEM) and a FEI Titan[3] 80-300 FEG-TEM, both equipped with probe and imaging aberration correctors. Images were acquired at 300 kV, using an 18 mrad probe-forming aperture, and 39-200 mrad detector collection angle. Quantitative analyses of the location of intensity maxima in STEM-ADF images were performed using an open-source Python package Atomap[38], where both the precise position and shape of atomic columns are measured using 2D elliptical Gaussian distributions.

TEM specimens were transferred in a glove box from the light-proof, $N_2$-filled transfer unit and mounted on a single tilt holder into the vacuum of the TEM. All TEM experiments followed a "shoot blind" protocol whereby zero electron dose was applied to the region-of interest during the experiment set up, neither during zone axis tilting nor parameter tuning. Kinematic diffraction pattern simulations were performed using SingleCrystal software.

To obtain STEM-ADF images suitable for quantitative measurements while preserving the structural integrity of the material, the electron dose was minimized by using dose-fractionation: 10 frames were acquired, each with fluence calibrated using a direct-counting detector and set to 1500 e/Å$^2$. Frames were carefully inspected to confirm that no structural changes occurred and then summed up to increase the signal-to-noise ratio.

In-situ heating experiments were conducted using a Tecnai F20 FEG-TEM equipped with a Gatan furnace holder. To minimize electron dose and beam damage, the electron beam was blanked during temperature changes and un-blanked only when setting temperatures had been reached and stabilized for image acquisition. For in-situ movies, images were acquired continuously at a rate of 1 frame/s.

## ASSOCIATED CONTENT

### Supporting Information

This material is available free of charge via the Internet at http://pubs.acs.org. Additional TEM characterization of ferroelastic-domain scaling; simulated diffraction patterns; geometric-phase analysis; atomic-resolution characterization of twin and antiphase walls; STEM-ADF mistilt simulations; structural models; correlated Cs-displacement and octahedral-tilt analysis; and additional in situ heating data (PDF).

**Corresponding Author.**

joanne.etheridge@monash.edu

**Author Contributions**

J.E. designed the study and supervised the project. W.L. designed and carried out TEM experiments and analyzed data. Q.Y. optimized and prepared specimens under the supervision of M.B.J.. W.L. and J.E. prepared the manuscript. All authors contributed to the discussion of the results and revision of the manuscript.

**Funding Sources**

Financial support from the Australian Research Council (ARC) is appreciated. J.E. acknowledges ARC Discovery Project DP200103070 and ARC Laureate Fellowship FL220100202. M.B.J. acknowledges financial support from the Engineering and Physical Sciences Research Council (EPSRC) via EP/T025077/1, EP/X038777/1 and APP30818.

**Notes**

The authors declare no competing interests.

**Acknowledgements**

The authors acknowledge use of facilities within the Monash Centre for Electron Microscopy, a node of Microscopy Australia. The Thermo Fisher Scientific Spectra φ TEM was funded by ARC LE170100118 and the FEI Titan$^3$ 80-300 FEG-TEM was funded by ARC LE0454166.

**<u>Table of Contents</u>**

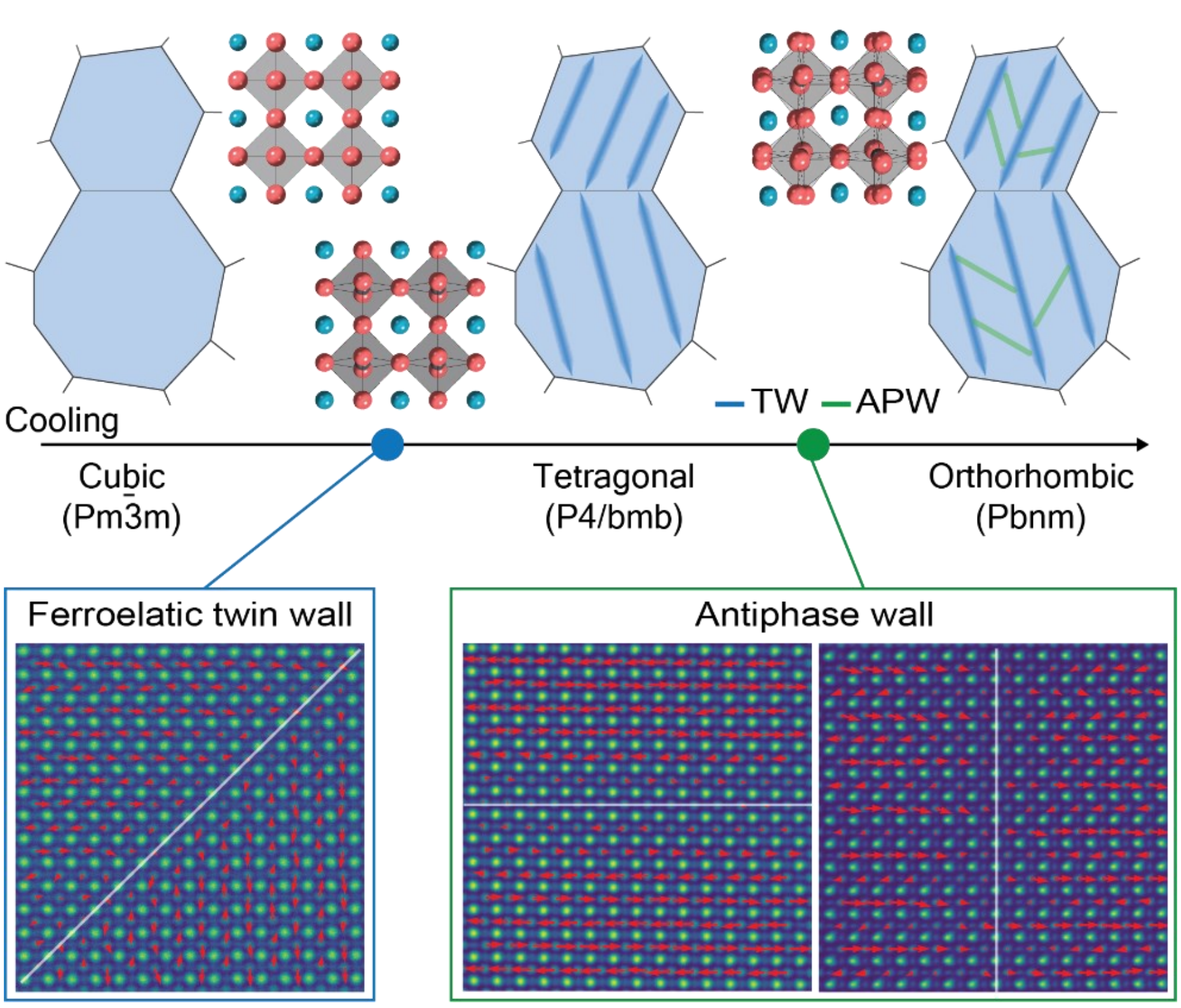